\documentclass[12pt,letterpaper]{article}

\usepackage{amsmath}
\usepackage{amsfonts}
\usepackage{amssymb}
\usepackage{graphicx}
\usepackage{psfrag}

\def\be{\begin{equation}}
\def\ee{\end{equation}}

\newcommand{\MP}{M_{P}}

\begin{document}

\begin{flushright} {\footnotesize IC/2007/001 \\ HUTP-07/A0002}  \end{flushright}
\vspace{5mm}
\vspace{0.5cm}
\begin{center}

\def\thefootnote{\fnsymbol{footnote}}

{\Large \bf A smooth bouncing cosmology\\[.1cm]
\bf with scale invariant spectrum\\[1cm]}
{\large Paolo Creminelli$^{\rm a}$ and Leonardo Senatore$^{\rm b}$}
\\[0.5cm]

{\small
\textit{$^{\rm a}$ Abdus Salam International Center for Theoretical
Physics\\ Strada Costiera 11, 34014 Trieste, Italy}}

\vspace{.2cm}

{\small
\textit{$^{\rm b}$ Jefferson Physical Laboratory \\
Harvard University, Cambridge, MA 02138, USA}}

\end{center}

\vspace{.8cm}

\hrule \vspace{0.3cm}
{\small  \noindent \textbf{Abstract} \\[0.3cm]
\noindent
We present a bouncing cosmology which evolves from the contracting
to the expanding phase in a smooth way, without developing
instabilities or pathologies and remaining in the regime of validity of
4d effective field theory. A nearly scale invariant spectrum of
perturbations is generated during the contracting phase by an
isocurvature scalar with a negative exponential potential and then
converted to adiabatic. The model
predicts a slightly blue spectrum, $n_S \gtrsim 1$, no observable gravitational waves and a high
(but model dependent) level of non-Gaussianities with local
shape. The model
represents an explicit and predictive alternative to inflation,
although, at present, it is clearly less compelling. \vspace{0.5cm}  \hrule
\def\thefootnote{\arabic{footnote}}
\setcounter{footnote}{0}

\section{Introduction}
Bouncing cosmologies, in which the present era of expansion is
preceeded by a contracting phase, have been studied as potential
alternatives to inflation in solving the problems of standard FRW
cosmology. The cosmological history is continued to the infinite past,
so that all the problems related to imposing extremely finely tuned initial
conditions in the high curvature regime are avoided. The pre Big-Bang
scenario \cite{Gasperini:2002bn} and the ekpyrotic/cyclic models
are different incarnations of these ideas \cite{Khoury:2001wf,Steinhardt:2001st}.

So far, two main problems have prevented these models from becoming
serious contenders to inflation.
\begin{enumerate}
\item No completely explicit model of a bouncing phase has been
  found. Both in the pre Big-Bang scenario and in the ekpyrotic/cyclic
  one the bouncing phase lies outside the regime of validity of the
  effective field theory, so that the issue requires some input from a
  more fundamental theory as string theory. The study of
  time-dependent solutions in string theory and of the way singularities are resolved
  is not sufficiently developed to describe a bouncing phase or even to
  assess if a bounce is possible altogether \cite{Gasperini:2002bn,Horowitz:2002mw,Turok:2004gb,Polchinski:2005qi,Tolley:2005us,Malkiewicz:2005ii,Niz:2006ef,Craps:2007iu}.

\item No compelling mechanism for producing an approximately
  scale-invariant spectrum of adiabatic perturbations has been
  found. The pre Big-Bang model predicted a dominant isocurvature
  mode, which is nowadays vastly incompatible with data. Although
  some alternatives have been studied, scale invariance does not come
  out naturally \cite{Gasperini:2002bn}.
  In the ekpyrotic/cyclic case it was argued that the same
  field leading to the fast contraction towards the bounce gives rise,
  with its quantum fluctuations, to an adiabatic, approximately scale invariant
  spectrum.  The issue is not completely straightforward because, as
  discussed, the bounce is not under full control so that the way
  perturbations evolve through it cannot be made explicit. The
  presence of a dynamical attractor simplifies the problem and
  predictions can be made under very general assumptions about the
  bounce \cite{Creminelli:2004jg}. Unfortunately the result is not
  scale-invariant \cite{Lyth:2001pf,Brandenberger:2001bs,Hwang:2001ga,Creminelli:2004jg}. One cannot
  rule out the possibility that the bouncing phase has very peculiar
  features which change the result \cite{Tolley:2003nx}, but in this case all predictions
  will depend on the details of the unknown phase.
\end{enumerate}

Recently there has been progress on both these issues and the purpose
of this paper is to put these new ingredients together to obtain an
explicit and controllable bouncing model with an approximately scale
invariant spectrum of density perturbations.
Let us discuss how to address the two problems discussed above.

\begin{enumerate}
\item In a recent paper \cite{Creminelli:2006xe} we studied explicit 4d models of a bouncing
  phase of the Universe. To induce the bounce, the stress-energy tensor
  must violate the null energy condition. In a spatially flat
  FRW metric this condition corresponds to the inequality $\dot H <0$,
  which is clearly violated at the point where we reverse from contraction to
  expansion: $H=0$ and $\dot H>0$. The violation of the null energy condition is usually
  associated with catastrophic instabilities which make the theory
  pathological. However this is not true in complete generality, as it
  can be shown with the quite general low energy effective field theory approach
  developed in \cite{Creminelli:2006xe}.
  An explicit realization can be obtained starting from the ghost
  condensate model \cite{Arkani-Hamed:2003uy} that we are now going to
  briefly describe. In an expanding
  Universe one expects that a  scalar field $\phi(t)$ evolving in time is progressively
  driven to rest by the Hubble friction. However, with a
  generic Lagrangian invariant under the shift symmetry $\phi \to \phi
  + {\rm const.}$
  \be
  {\cal{L}} = P\big((\partial\phi)^2\big) \;,
  \ee
   it is
  easy to realize that another possibility is that $\dot\phi$ goes to
  a constant, $\dot\phi \to M_{\rm gc}^2$, in correspondence of a minimum of the function $P$. The peculiarity of this solution is
  that, although the scalar keeps on evolving at constant speed, the
  stress energy tensor is the one of a cosmological constant, so that
  the background metric can be de Sitter or Minkowski. Fluctuations around
  this background solution are healthy and this theory has
  been studied as a consistent modification of gravity in the IR. To
  understand the relevance of this model in our context one can
  study the stress-energy tensor of the perturbations around the
  background solution: $\phi(\vec x,t) = M_{\rm gc}^2 (t + \pi(\vec
  x,t))$. $T_{\mu\nu}$ starts linearly in $\pi$ with a
  term \cite{Arkani-Hamed:2003uy}
\be
T_{\mu\nu} = M_{\rm gc}^4 \dot\pi \;\delta_{\mu 0} \delta_{\nu 0}\;.
\ee
This shows that a fluctuation with $\dot\pi <0$ has negative energy!
This property opens up the possibility of violating the null energy
condition. In particular an explicit model of a bounce can be obtained
adding to the Lagrangain above a suitable potential $V(\phi)$. We will
see for example that a bouncing phase with a constant positive $\dot
H$ can be obtained simply with a parabolic potential. In \cite{Creminelli:2006xe} we
checked that, with a suitable choice of the model parameters, no
dangerous instabilities are present.

Besides giving an existence proof of a smooth bounce, our results
allow to study the evolution of perturbations across this
phase. Nothing exotic happens in these models, so that the general
conclusions of \cite{Creminelli:2004jg} apply. Therefore the simple mechanism for producing a
scale invariant spectrum of perturbations proposed in the
ekpyrotic/cyclic scenarios does not work. The fluctuations in the
field leading the contracting phase have a very blue spectrum, which
is completely negligible on the scales of cosmological interest. This
motivates us to look for an additional source of perturbations.

\item Recently Lehners, McFadden, Steinhardt and Turok \cite{turok} pointed out that a second scalar
  field whose energy density is negligible and that evolves along a negative
  exponential potential could be the source of a
  scale-invariant spectrum of perturbations. These fluctuations can be easily
  converted to adiabatic and therefore match observations. Early
  studies of isocurvature perturbations in bouncing scenarios can be
  found in \cite{Notari:2002yc,Finelli:2002we}.
\end{enumerate}

A possible emerging picture is therefore as follows. The Universe is
contracting and its energy density is dominated by a field $\phi$,
with a Lagrangian similar to the ghost condensate theory, with the
addition of a suitable potential. As in the ekpyrotic/cyclic scenario
the contraction satisfies $\dot H \ll -H^2$. At a certain point $\dot H$
flips sign until $H$ gets to zero and the Universe starts expanding
and smoothly connects, after reheating, to a standard FRW cosmology. During the
contracting phase a second field $\psi$ evolves along a negative
exponential potential. Perturbations in this ``isocurvature'' field
are then converted into adiabatic ones: as $\phi$ perturbations are
very suppressed, this second source gives the leading contribution to
cosmological inhomogeneities.

We are going to study the model starting in Section \ref{sec:exp} from the field $\psi$, which
is the source of perturbations and therefore the sector of the model
giving all the predictions. We will see that, neglecting gravity, an
exactly scale-invariant spectrum of perturbations is obtained for any
negative exponential potential. Gravity
slightly modifies the spectrum and {\em the resulting tilt is blue}. The
deviation from scale invariance may be very small, but it is not
possible to get a red spectrum. This prediction is disfavored by
recent data \cite{Spergel:2006hy} and the model will therefore be ruled out in the very near
future if the detection of a red tilt is confirmed. Additional
predictions are the presence of a quite high (but model dependent)
level of non-Gaussianity (with a local shape dependence) and an
unobservably small level of primordial gravitational waves (as in the
ekpyrotic/cyclic scenarios).
In Section \ref{sec:story} we study the unperturbed evolution led by the
$\phi$ field. It is important to stress that all the details are very
model dependent and that even the use of a ghost-condensate like
theory should be considered simply as a working example of an healthy
bouncing cosmology. As all predictions do not depend on how the
contracting and 
bouncing phases are realized, one can envisage other very different possibilities.
Some remarks about
the exponential form of the potential, which is the root of scale invariance in our model, is made in Section
\ref{sec:radiative} before the conclusions in Section
\ref{sec:conclusions}. In the Appendix \ref{app: full action} we study
the full Lagrangian of perturbations of $\phi$ and $\psi$, including
mixing terms.

\section{\label{sec:exp} Exponential is good}
Let us for the moment completely neglect gravity. Consider a scalar
$\psi$ with standard kinetic term and potential $V(\psi)$. The unperturbed
space independent solution $\psi(t)$ will satisfy
\be
\label{eq:Newton}
\ddot\psi(t) + V'(\psi(t)) =0 \;.
\ee
Perturbations around it will be denoted by $\delta\psi(\vec x, t)$. We want to answer the following question: which form of the function
$V(\psi)$ gives rise to a scale invariant spectrum of perturbations
$\delta\psi$?

The equation of motion for $\delta\psi$ is
\be
\label{eq:deltapsi}
\delta\ddot\psi + \left[k^2 + V''(\psi(t))\right] \delta \psi =0 \;.
\ee

If $V''<0$, when the gradient term becomes negligible
with respect to the mass term the mode ``freezes'', {\em i.e.} the
solutions of the equation above are no longer oscillatory.
To have a power-law solution for $\delta\psi(t)$ in this
regime one requires $V''(\psi(t)) \propto t^{-2}$. For $t<0$ the mass term increases (in modulus) and more and more modes freeze.

As we are assuming $V''(\psi(t)) \propto t^{-2}$, equation
(\ref{eq:deltapsi}) has solutions of the form
\be
\psi_k \sim \frac1{\sqrt{k}} \cdot F(kt)
\ee
in the conventional normalization.

Therefore to get a scale invariant spectrum the evolution after freezing must be of the form
$\delta\psi \propto t^{-1}$.
This time evolution is solution only if
\be
\label{eq:mass}
V''(\psi(t)) = -\frac2{t^2} \;.
\ee
With this choice, besides the solution $\delta\psi \sim t^{-1}$, we
have a decaying solution $\delta\psi \sim t^2$. The explicit solution
with the standard normalization is given by
\be
\delta\psi_{\vec k} =\frac1{\sqrt{2 k}} e^{-i k t} \left(1-
  \frac{i}{kt}\right) \;.
\ee
Clearly the result of eq.~(\ref{eq:mass})
could have been anticipated, as eq.~(\ref{eq:deltapsi}) is now the
same as the one describing a massless scalar living in de-Sitter space
(replacing $t$ with the conformal time $\eta$).

Now we should look at eq.~(\ref{eq:mass}) as an equation for the
function $V(\psi)$. Taking the time derivative of equation
(\ref{eq:Newton}) we get
\be
\dddot\psi -\frac2{t^2} \dot\psi =0 \;.
\ee
Neglecting a decaying term this gives $\dot\psi(t) =-2 M/t$,
with $M$ an integration constant with dimension of a mass. Thus we
have
\be
\psi(t) = - 2 M \log(-t) + c \;.
\ee
We can now get rid of time in the second derivative of the potential $V''(\psi(t)) = -\frac2{t^2}$
and then integrate to finally get
\be
V(\psi) = - V_0 \; e^{\psi/M} \;.
\ee


We thus reach the conclusion that, in the absence of gravity, a field moving along any negative exponential potential
generates a scale invariant spectrum of perturbations
\cite{Khoury:2001zk}. Notice that while generating say 60 e-folds of
scale invariant spectrum the potential varies by $e^{60}$. This
implies that we really need an exponential potential, and that our
result is not an artifact of requiring {\em exact} scale
invariance\footnote{Actually we do not know if the
  perturbations are scale invariant over 60 e-folds, but only in a
  much narrower range. For sure we need an exponential potential in
  the observable window of $\sim 10$ e-folds.}.

We now want to see whether this way of generating a flat spectrum
survives when we put gravity back into the game. If the field
$\psi$ is cosmologically relevant, {\em i.e.} if its contribution
to the total stress-energy tensor is significant, then the picture
drastically changes due to its mixing with gravity
\cite{Creminelli:2004jg}. Therefore we are going to assume on the
contrary that the energy of $\psi$ is negligible so that we can
forget about its mixing with gravity \footnote{The precise regime
in which this approximation is justified depends not only on
the relative contribution to the energy density, but also on the
nature of the field that drives the expansion of the universe. In
Appendix \ref{app: full action} we shall determine precisely the
conditions for our approximation to be valid; for the moment we
just assume that we are in such a
regime.}.

Let us assume that $\psi$ lives in a contracting Universe with $a
\propto |t|^p$, $H = \frac{p}{t}$, with $t<0$ and $p
< 1/3$. When this last inequality is satisfied, the background solution blue-shifts so fast that gets rid of
initial curvature, inhomogeneities and anisotropies. In this background the equation of motion for $\psi(t)$ is given by
\be
\label{eq:psiwithH}
\ddot\psi +\frac{3p}{t}\dot\psi -\frac{V_0}{M}e^{\psi/M} =0 \;.
\ee
It is straightforward to check that we still have a power-law solution with
\be
\label{eq:solp}
\dot\psi = -\frac{2M}{t} \qquad V = - \frac{2M^2(1-3p)}{t^2} \;.
\ee
The total energy of $\psi$ is obviously no longer conserved and it is
given by $\rho_\psi=6pM^2 t^{-2}$.  This contribution will be
negligible with respect to the total energy density of the
Universe $\rho = 3 M_P^2 H^2 = 3 p^2 M_P^2 t^{-2}$, for sufficiently
small $M$. The solution described above is tuned, as the divergence of
$\psi$ and $H$ both happen at $t=0$. This case is clearly not generic,
but if we shift the background solution $H=p\,(t+t_0)^{-1}$, we see
that eq.~(\ref{eq:solp}) is still an approximate solution of
(\ref{eq:psiwithH}) going sufficiently back into the past: $|t| \gg
|t_0|$.  Let us study the behaviour of perturbations, that
are now going to deviate from an exactly scale invariant spectrum.

{\bf Tilt of the spectrum}.
In the presence of gravity, the equation describing the perturbations is
\be
\delta\ddot\psi +\frac{3p}{t} \delta\dot\psi + \left[\frac{k^2}{t^{2
    p}}-\frac{2(1-3p)}{t^2}\right] \delta \psi =0 \;.
\ee
The equation can be simplified using the conformal time $\eta$ and the variable
$u = a \delta\psi$. At first order in $p$ we obtain
\be
\frac{d^2 u}{d \eta^2} + (k^2 - \frac{2-3p}{\eta^2}) u = 0 \;.
\ee
The solution of this equation with the correct limit at short distance
is given by
\be
u_k(\eta) = \frac{\sqrt\pi}{2} e^{i \frac\pi4 (1 + \sqrt{9-12 p})}
\sqrt{-\eta} \, H_{\frac12 \sqrt{9-12 p}} \; (-k\eta) \;.
\ee
Taking the limit $\eta \to 0$ we get that the function goes, at first
order in $p$, as $k^{-3/2+p}$. The tilt is thus blue and given by
\be
n_s-1 = 2 p \;.
\ee
As there is no natural, smooth transition between the rapidly
contracting phase when perturbations are produced and the bounce, we
cannot establish a ``natural'' value for $n_s - 1$; the deviation from
scale invariance can be arbitrarily tiny. On the contrary
it is quite natural in inflationary models to have deviations from
scale invariance (of both red and blue kind) of order $n_s-1 \sim \pm \,
N_e^{-1} \sim \pm \,{\rm few} \,\%$, where $N_e$ is the number of e-folds to the end of inflation. Moreover here we
do not have any analogue of the slow-roll expansion, so that there is
no reason why the tilt should be approximately a constant or
equivalently why the parameter $p$ should not change in
time\footnote{In the text we consider the case in which the ratio
  $H^2/|V''(\psi)|$ is small and constant, which is analytically
  simpler. However it seems rather tuned to assume that the two quantities $H^2$ and
  $V''$ vary by many orders of magnitude keeping their ratio
  constant. To have approximate scale invariance we just need this
  ratio to remain small, though it can vary significantly while modes
  of cosmological interest freeze.}: in particular the tight constraints that we
have on $p$ on large scales cannot be extrapolated to
short scales.  The only sharp prediction is that
the tilt {\em cannot be red}.


Fluctuations in the $\psi$ field will be
converted into adiabatic perturbations, which can be described by the
usual variable $\zeta$. We postpone to the next Section the discussion
about possible conversion mechanisms. The conversion will happen when
modes of cosmological interest are frozen and gradients are
irrelevant, so that the tilt of the spctrum cannot change.
At linear order
\be
\zeta = \frac{\delta\psi}{M_c} \simeq \frac{t_0^{-1}}{M_c} \;,
\ee
where $M_c$ is a parameter with dimension of a mass, which depends
on the conversion mechanism.

Let us now look at the other observables.

{\bf Gravity waves}. Relic gravitational waves will have a very blue
spectrum since they are just sensitive to the ratio $H/M_P$ during the
contracting phase and will be therefore strongly suppressed on large
scales. The situation is very close to ekpyrotic/cyclic models
\cite{Boyle:2003km} and no detection is possible in the foreseeable future.

{\bf Non-Gaussianities}. There are two sources of non-Gaussianity one
has to consider. First of all, the exponential potential is non-linear
and therefore it sources interactions among different $\delta\psi$ modes. This is similar to what happens in inflationary models with a
second field: self-interactions of this isocurvature scalar
induce a non-linear evolution, so that the
statistics of the field becomes non-Gaussian. The level of non-Gaussianity, {\em i.e.}~the
correction to the linear theory, can be obtained comparing the mass
term $\frac12 V''(\psi) \delta\psi^2$ with the cubic term $\frac{1}{3!}V'''(\psi)
\delta\psi^3$. This gives
\be
{\rm NG} \simeq \frac{\delta\psi}{M} \simeq  \frac{|t|^{-1}}{M} \;.
\ee
As $\delta\psi$ increases with time, most of the non-linearity will
develop at the very end. The resulting non-Gaussianity is thus of
order $1/(|t_0| M)$, where $t_0$ is the latest time where the solution
described in the previous Section applies.

The leading non-Gaussianity is cubic (quartic and higher terms are
suppressed with respect to the cubic by further powers of
$\delta\psi/M$) and it will show up as a 3-point function
correlator. As non-linearities are dominated by the late time
evolution, when gradient terms can be neglected, the momentum
dependence of the 3-point function will be of the ``local'' form.
The same happens in multi-field inflationary models where the out
of the horizon evolution dominates the non-Gaussianity
\cite{Zaldarriaga:2003my,Babich:2004gb}.

The explicit calculation, neglecting ${\cal O}(p)$ corrections, is
quite simple. Following Maldacena \cite{Maldacena:2002vr} the 3-point
correlator at the final time $t_0$ is given at tree level
by\footnote{The tree level calculation can be interpreted as the
  classical non-linearity among the modes generated by the
  background. There will be quantum corrections to this calculation
  coming from loop diagrams and thus suppressed by higher powers of
  $\hbar$ \cite{Weinberg:2005vy}. In our case an interaction of the
  form $V^{(n)}\delta\psi^n/n!$ will give a correction to the 3-point function
  of order $1/(t M)^2 (E/M)^{n-4}$, where $E$ is the
  typical energy of the process $E \sim |t|^{-1}$. The cubic term is
  the most important giving a correction $\sim 1/(|t|M)$, which is
  anyway small as the tree level calculation forces $1/(|t_0| M) \ll 1$.}
\be
\langle\delta\psi_{\vec k_1}(t_0) \delta\psi_{\vec k_2}(t_0) \delta\psi_{\vec k_3}(t_0) \rangle = - i
\int_{-\infty+i \epsilon}^{t_0} \langle \delta\psi_{\vec k_1}(t_0)
\delta\psi_{\vec k_2}(t_0) \delta\psi_{\vec k_3}(t_0) \; H_{\rm int}(s) \rangle ds +  c.c.
\ee
where the interaction Hamiltonian is the cubic $\delta\psi$ self-interaction
\be
H_{\rm int}(t) = \frac{V'''(t)}{3!} \delta\psi^3 = -\frac2{M t^2} \frac{\delta\psi^3}{3!}  \;.
\ee
Using the normalized free field solution
\be
\delta\psi_{\vec k} =\frac1{\sqrt{2 k}} e^{-i k t} \left(1- \frac{i}{kt}\right)
\ee
we get
\begin{eqnarray}
\nonumber
\langle\delta\psi_{\vec k_1}(t_0) \delta\psi_{\vec k_2}(t_0)
\delta\psi_{\vec k_3}(t_0) \rangle & \!\!\!\!\!= &\!\!\!\!\!(2 \pi)^3 \delta(\sum_i \vec
k_i) \frac1{\prod_i 2 k_i^3} \left(\frac1{t_0}+i
  k_1\right)\!\left(\frac1{t_0}+i k_2\right) \!\left(\frac1{t_0}+i
  k_3\right) e^{-i (k_1+k_2+k_3) t_0}  \\ & & \!\!\!\!\!\!\!\!\!\!\!\!\!\!\!\!\!\!\!\!\!\!\!\!\!\!\!\int_{-\infty+i
  \epsilon}^{t_0} \left(k_1+\frac{i}{s}\right)
\left(k_2+\frac{i}{s}\right) \left(k_3+\frac{i}{s}\right) \cdot
e^{i (k_1+k_2+k_3) s} \left(-\frac{2}{M s^2}\right) ds + c.c.
\end{eqnarray}
We are interested in the leading contribution for small $t_0$, which is
given by
\be
\langle\delta\psi_{\vec k_1}(t_0) \delta\psi_{\vec k_2}(t_0) \delta\psi_{\vec k_3}(t_0) \rangle = (2 \pi)^3 \delta(\sum_i \vec k_i) \frac{\sum_i k_i^3}{\prod_i k_i^3} \cdot \frac1{8 |t_0|^3} \frac{1}{M |t_0|} \;.
\ee
We recognize in front the expected ``local" momentum dependence. With
the usual definition of the power spectrum of $\delta\psi$ at time
$t_0$: $\langle\delta\psi_{\vec k_1} \delta\psi_{\vec k_2}\rangle
\equiv (2\pi)^3 \delta(\vec k_1 + \vec k_2) k_1^{-3} \cdot \Delta_\psi$, $\Delta_\psi=1/(2 t_0^2)$, we can rewrite the result as
\be
\langle\delta\psi_{\vec k_1}(t_0) \delta\psi_{\vec k_2}(t_0) \delta\psi_{\vec k_3}(t_0) \rangle = (2 \pi)^3 \delta(\sum_i \vec k_i) \frac{\sum_i k_i^3}{\prod_i k_i^3} \Delta_\psi^{3/2}  \cdot \frac{\sqrt{2}}{4} \frac1{M |t_0|} \;.
\ee
We see explicitly that the non-linear corrections are of order $1/(M |t_0|)$ as discussed above.

The local form of non-Gaussianity is usually defined through the relation
\begin{equation}
\label{eq:f_NL}
\zeta(x) =\zeta_g(x) -\frac35 f^{\rm local}_{\rm NL}(\zeta_g(x)^2 - \left<\zeta_g^2\right>) \;,
\end{equation}
where $\zeta$ is the observed perturbation and $\zeta_g$ is a Gaussian
variable. Experimental limits are given in
terms of the scalar variable $f_{\rm NL}^{\rm local}$. The variable
$\zeta$ will be related to $\delta\psi$ as $\zeta(\vec x)= \pm
\delta\psi(\vec x)/M_c$, with $M_c$ a mass scale depending on the
conversion mechanism. Notice that we have a sign ambiguity: depending
on the mechanism, positive $\delta\psi$ will correspond to
positive $\zeta$ or viceversa. Possible additional non-linearities in this relationship will be discussed below.

The 3-point function of $\zeta$ will thus be given by
\be
\langle\zeta_{\vec k_1} \zeta_{\vec k_2} \zeta_{\vec k_3} \rangle =
\pm (2 \pi)^3 \delta(\sum_i \vec k_i) \frac{\sum_i k_i^3}{\prod_i k_i^3} \Delta_\zeta^{3/2}  \cdot \frac{\sqrt{2}}{4} \frac1{M |t_0|} \;,
\ee
which implies
\be
\label{eq:locexp}
f^{\rm local}_{\rm NL} = \mp \frac{5 \sqrt{2}}{24} \Delta_\zeta^{-1/2} \frac1{M |t_0|} \;.
\ee

The tightest experimental constraints on $f_{\rm NL}^{\rm local}$ are presently
coming from the analysis of WMAP 3yr data \cite{Spergel:2006hy, Creminelli:2006rz}
\be
\label{eq:locfinal}
-36 <f_{\rm NL}^{\rm local} < 100 \quad {\rm at} \;95\% \;{\rm C.L.}
\ee
This constraint gives
\be
\frac1{M |t_0|} < 100 \cdot \frac{24}{5 \sqrt{2}} \,\Delta_\zeta^{1/2} \simeq 6 \cdot 10^{-2} \quad  {\rm at} \;95\% \;{\rm C.L.}
\ee

The second source of non-Gaussianity is the conversion between isocurvature perturbations $\delta\psi$ and the adiabatic mode $\zeta$. As everything happens when gradients are irrelevant the relationship between the two variables will be local in real space, but in general not linear
\be
\zeta(\vec x)= \delta\psi(\vec x)/M_c + f_{\rm NL}^{\rm local}  (\delta\psi(\vec x)/M_c)^2  + \ldots
\ee
The quadratic correction gives an additional contribution to the local non-Gaussianity calculated in eq.~(\ref{eq:locexp}). What is experimentally constrained is the sum of the two parameters $f_{\rm NL}^{\rm local}$. The contribution from the conversion mechanism cannot be further studied without specifying the model; only the local shape dependence of the 3-point function is model independent.

In this Section we have derived all the predictions assuming that the
energy density of the field $\psi$ is so small that its mixing with
gravity can be neglected, at least when modes relevant for cosmology freeze. In Appendix \ref{app: full action} we study
the full action describing both $\psi$ and the field $\phi$, which dominates
the stress energy tensor. In this way we can check under which
conditions the mixing of $\psi$ with gravity can be neglected. If the
mixing cannot be disregarded, predictions will change and become more
model dependent, as they will depend on the Lagrangian of $\phi$. In
this case our sharp prediction of a blue tilt might also change. We leave these scenarios for
future study.

\section{\label{sec:story}The whole story}
Now that we have discussed the dynamics of perturbations sourced by
the $\psi$ field, let us concentrate on the unperturbed history of our
bouncing cosmology. We will assume that the entire history, until the
beginning of the standard hot FRW era, is dominated by the scalar
$\phi$, with the Lagrangian of a ghost condensate theory with the
addition of a suitable potential.
As it should be clear from the discussion above,
this model should be regarded as an explicit example of a smooth
evolution from contraction to the bounce and then to the expanding
phase, which is not affected by fast instabilities or other pathologies. As predictions do not depend on the explicit
realization of the unperturbed history, alternative models may be
found, either at the level of field
theory or in which stringy effects are important.

\subsection{Contracting phase}
Although not required, we saw
in the previous Section that it is
simpler to study an evolution with $a \sim |t|^p$ with $p$
constant. Furthermore we demand $p \ll 1$, so that the spectrum is
close to scale invariance.
We want to find out which form of the potential $V(\phi)$ must be added to the
ghost condensate Lagrangian to obtain this evolution of the scale
factor. Neglecting terms with more than one derivative acting on
$\phi$ we have
\be
{\cal L} = \sqrt{-g} \left[P(X,\phi) - V(\phi)\right] \qquad X \equiv
- g^{\mu\nu} \partial_\mu\phi\partial_\nu\phi \;.
\ee
Notice that we are discussing a slightly more general situation with
respect to the Introduction, in which the shift
symmetry of $\phi$ is not only broken by the potential, but also in
the generalized kinetic term $P$. The reason for this will become
clear later on. For concretness we can take a functional form
$P(X,\phi) = \phi^{-2} \bar P(X)$. In the absence of a potential, it
is easy to check that,
if $\bar P$ has a minimum where it vanishes ($\bar
P'(\dot\phi_0^2) = \bar P(\dot\phi_0^2)=0$), we have the solution $\phi(t)
= \dot\phi_0 t$ with a Minkowski metric (or de Sitter in the
presence of a positive cosmological constant). In this case the
system behaves as the original ghost condensate, with the only
difference that, as $P_{,XX}(\dot\phi_0^2,\phi_0(t)) = (\dot\phi_0
t)^{-2} \bar P''(\dot\phi_0^2)$ depends on time, the dynamics of
perturbations progressively changes.

The stress energy tensor is given by
\begin{equation}
T_{\mu\nu} = [P(X,\phi) - V(\phi) ]g_{\mu\nu} + 2 P(X,\phi)_{,X}
\partial_\mu\phi\partial_\nu\phi \;.
\end{equation}
For sufficiently small departure from $\bar{P}'=0$, the part of
the stress energy tensor that depends on the generalized kinetic
term is linear in $\dot\pi$ and it contributes only to the energy
density, without pressure. In the Friedmann equation for $\dot H$
the contribution of the potential cancels: \be \label{eq:dotH}
M_P^2 \dot H = -\frac12 (\rho+ p) = -\frac12 \cdot 4
P_{,XX}\dot\phi_0^4 \,\dot\pi = -\frac12 \frac{\tilde M_{\rm gc}^2}{t^2}
\dot\pi \ ,\qquad \tilde M_{\rm gc}^2 \equiv 4 \bar P'' \dot\phi_0^2\;. \ee
In the last equality we have assumed that the deviation from the
solution in the absence of potential, $\phi(t) = \dot \phi_0 t$,
is small. Assuming this, we can solve for $\pi$ as a function of
$t$: \be \label{eq:dotpi} -M_P^2 \frac{p}{t^2} =  -\frac12
\frac{\tilde M_{\rm gc}^2}{t^2} \dot\pi \quad \Rightarrow\quad \dot\pi =
\frac{2 p \, M_P^2}{\tilde M_{\rm gc}^2} \;. \ee The velocity of $\phi$ is
shifted by a constant amount. The deviation from the solution in
the absence of a potential is therefore small if $\dot\pi=\frac{2
p \, M_P^2}{\tilde M_{\rm gc}^2}\ll 1$, as we are going to assume in the
following. We can now solve the first Friedmann equation to find
$V(\phi)$ \be V(t) = 3 M_P^2 H^2 +2 M_P^2 \dot H = \frac{M_P^2 (3
p^2- 2p)}{t^2} \;. \ee In the approximation $\dot\pi \ll 1$ we
have $\phi(t) = \dot\phi_0 t$ so that the potential turns out to
be \be V = -\frac{(2p-3p^2)M_P^2 \dot\phi_0^2}{\phi^2} + {\cal
O}(\dot\pi)\;. \ee We are interested in the regime $p \ll 1$ so
that we obtain an inverse quadratic potential which is negative
and going to $-\infty$ for $t \to 0$. This solution makes sense.
In the absence of a potential, starting from $\dot \pi >0$, we
would have a matter-domination like solution, as the contribution
to $T_{\mu\nu}$ proportional to $\dot\pi$ does not have pressure.
This would correspond to $p = \frac{2}{3}$ and in fact $V=0$ in
the equation above for $p=2/3$. As $t^{-2 }\dot\pi \propto \dot
H\propto a^{-2/p}$, for smaller $p$ we need $t^{-2}\dot\pi$ to
increase faster as a function of the scale factor and this
corresponds to a negative tilted potential which induces an
increase in the $\phi$ velocity. Viceversa for $p > \frac23$.

It is important to stress that the solution above is a dynamical
attractor in the same way as in the ekpyrotic/cyclic models. One
can check that if one starts with a $\dot\pi$ slightly different
from the one of eq.~(\ref{eq:dotpi}), this perturbation to the
velocity dies as $t$ and the solution goes back to the unperturbed
one. Also curvature and anisotropies become irrelevant, as the
energy density increases with the scale factor faster than
$a^{-6}$ \cite{Erickson:2003zm}. For a local observer the
evolution converges to a single unperturbed history, which turns
out to be a crucial simplification to allow following
perturbations across the bounce \cite{Creminelli:2004jg}. This
last point is not so relevant in our case as we have an explicit
bounce and we can follow the perturbations throughout.



To conclude this Section we are now going to study perturbations of
$\phi$; the purpose is just to check that it is consistent to neglect
them and to have only the $\psi$ sector as source of the observed
perturbations.

Following the general results of \cite{Creminelli:2006xe} in Appendix
\ref{app: full action} we derive the Lagrangian for the usual variable
$\zeta$ which describes the scalar
perturbations of the system\footnote{What is relevant for observations
  is the constant mode of $\zeta$ \cite{Creminelli:2004jg}. As we are interested in studying
  the behaviour of $\phi$ perturbations, we simply disregard the
  $\psi$ field in this Section.}. We simply get
\begin{equation}\label{zetalag}
S = \int d^3 x\; dt\; a^3(t) \left[\frac12 \frac{\tilde M_{\rm gc}^2
t^{-2}}{H^2}
  \dot\zeta^2 - \frac{M_P^2}{p} \left(\frac{\partial}{a}\zeta\right)^2\right] \;.
\end{equation}
One can show by symmetry arguments  \cite{Creminelli:2006xe} that the coefficient of
the spatial kinetic term $(\partial \zeta/a )^2$ is always given by $M_P^2 \dot H/H^2$,
independently of the $\phi$ Lagrangian.
The time kinetic term is not fixed by symmetries and in our case it is
time-independent; therefore $\zeta$ modes
have a constant speed of sound $c_s=\sqrt{2p} \,M_P/\tilde M_{\rm gc} \ll 1$. The
amplitude of a $\zeta$ mode at freezing will be
\begin{equation}
\zeta \sim H \cdot \frac{H}{\tilde M_{\rm gc} t^{-1}} \frac{1}{c_s^{3/2}}
\;,
\end{equation}
where $t$ and $H$ must be evaluated at horizon crossing. The
amplitude grows as $H$ so that the spectrum is strongly blue.
Neglecting $p$-corrections $H_{\rm freezing}(k) \propto k$, so
that the final spectrum goes as $1/k$ instead of the usual
$k^{-3}$ for the scale invariant case. This also tells us that the
$\zeta$ perturbations are completely negligible on scales of
cosmological interest.

As we discussed, in the limit $\dot H \to 0$ the spatial kinetic
term in (\ref{zetalag}) vanishes. For this reason it is important
to consider higher derivative terms which in the $\pi$ Lagrangian
give \cite{Arkani-Hamed:2003uy,Creminelli:2006xe}
\begin{equation}
-\frac12 \bar M_{\rm gc}^2 (\partial^2\pi/a^2)^2 \;.
\end{equation}
This term is the leading spatial kinetic term in the original
ghost condensate model for which $\dot H =0$ and it is very
important to control the stability of the system as we will
discuss in the next Section.

However, as a consequence of the mixing with gravity, the higher
derivative term proportional to $\bar M_{\rm gc}^2$ induces an
additional 2-derivative term in eq.~(\ref{zetalag}) of the form
\be \frac{\bar M_{\rm gc}^2 \tilde{M_{\rm gc}}^2t^{-2}}{2 H^2 M^2_P}
\left(\frac{\partial}{a} \zeta\right)^2 \ee which has the wrong,
unstable sign and describes a Jeans-like instability
\cite{Arkani-Hamed:2003uy}.
The system is stable if the term
$(\partial\zeta)^2$ in eq.~(\ref{zetalag}), which has the healthy
sign, dominates. This happens for
\begin{equation}
\bar M_{\rm gc}\ll c_s M_P \;.
\end{equation}
Restricting to this
regime, one can also see that around the time when a mode freezes
the term $(\partial^2\zeta/a^2)^2$ is completely negligible, and
this justifies our use of the simplified Lagrangian
(\ref{zetalag}) for the study of the fluctuations generated during
the contracting phase.


\subsection{The bounce}
The very same procedure can be used to get the potential for the
bouncing phase. For simplicity we come back to the usual ghost
condensate with a non standard kinetic term of the form $P(X)$,
with the shift symmetry broken only by a potential term. In this case
we can assume without loss of generality that $4 P'' \dot\phi^4 =
\dot\phi^2 \equiv M_{\rm gc}^4$.

As an example we take $H$ to be a linear function of time across
the bounce
\begin{equation}
H(t) = \frac{t}{T^2} \qquad \dot H = \frac1{T^2} \;.
\end{equation}
As $\dot H$ is constant eq.~(\ref{eq:dotH}) implies a constant
negative $\dot\pi$ so that \be \pi = - \frac{2 M_P^2}{M_{\rm gc}^4
T^2} t \;,
\end{equation}
where we choose the
integration constant such that $\pi$ vanishes when the Universe
reverses from contraction to expansion. In the limit $|\dot\pi|
\ll 1$, we get the parabolic potential \be V(\phi) = \frac{2
M_P^2}{T^2} + \frac{3 M_P^2 \phi^2}{(M_{\rm gc}
  T)^4} \;.
\end{equation}
We start in contraction with a positive potential energy
and a negative contribution from the $\dot \pi$. The potential
decreases in time, while  $\dot \pi$ stays constant until the
total energy density goes to zero and the Universe bounces to an
expanding phase.

The dispersion relation for the fluctuations of the ghost
condensate as calculated in \cite{Creminelli:2006xe} for a generic
FRW background is
\begin{equation}
\omega^2 = - \frac{\bar M_{\rm gc}^2 M_{\rm gc}^4 +4 M_P^4 \dot
H}{2M_P^2 M_{\rm gc}^4} k^2  + \frac{\bar M_{\rm gc}^2}{M_{\rm
gc}^4} k^4\;.
\end{equation}
As discussed in the last Section, there is a $k^4$ term proportional to
$\bar{M}_{\rm gc}^2$, and two terms going as $k^2$. The difference with
the last Section is that now both the $k^2$ terms have the unstable
sign. The one proportional to $\bar M_{\rm gc}^2$ describes the Jeans
instability which is a consequence of the mixing with gravity. The second term is proportional to $\dot
H$, as we remarked in the last Section, so that it has now the unstable sign.

It is straightforward to check \cite{Creminelli:2006xe} that one cannot make both these sources
of instability arbitrarily small: the best compromise, {\em i.e.}~the choice
which minimizes the instability rate of the system, is to make the two
instabilities comparable and this happens for $\bar M_{\rm gc}
M_{\rm gc}^2/M_P^2 \simeq \dot H^{1/2}$. In this case it is easy
to check that the most unstable mode has a rate of instability
$\omega_{\rm max} \simeq \dot H^{1/2} = \frac1{T}$. As the
duration of the bounce is of order $T$, this mode grows by order
one and this is clearly not problematic.

One may worry about the relevance of this
instability for predictions, as modes of cosmological
interest evolve during the bouncing phase.
Intuitively however one expects that if the bounce occurs
sufficiently fast, the time during which the very long wavelength
modes evolve is much
shorter than their typical frequency, making the effect of this
evolution irrelevant for cosmological observations. Let us check that this is indeed the case
and the effect is completely negligible. As one can easily verify,
the most unstable mode has a wavelength much smaller than
$H^{-1}(T)$, so that for modes of cosmological interest, which are
obviously with wavelengths much larger than $H^{-1}(T)$, one can
neglect the $k^4$ term and take the dispersion relation to be
\begin{equation}
\omega^2 = - \left(\frac{M_P}{M_{\rm gc}^2 T}\right)^2 k^2 \;.
\end{equation}
Each mode will grow by a factor $\omega(k) \cdot T$, where we have
taken the duration of the bounce to be of order $T$ . Assuming
that just after the end of the bounce we begin a hot FRW
cosmology, a given mode can be characterized as in inflationary
cosmology with its number of e-folds. In this context this just
tells us how much its wavelength is larger than the horizon at the
beginning of the FRW era. We obtain
\begin{equation}
\omega \cdot T = \frac{M_P}{M^2_{\rm gc}}\frac1{T}\; e^{- N}\sim
\frac{\bar M_{\rm gc}}{M_P} e^{-N} \;.
\end{equation}
where in the second passage we have taken the relation $T^{-1} =
\dot H^{1/2} = \bar M_{\rm gc} M_{\rm gc}^2/M_P^2$, the choice
which minimizes the instability rate. The effect is in any case
exponentially small and thus completely irrelevant for
observations.

The two phases described above, the contraction with constant $p$
and the bounce, must be somehow glued together. One can imagine
for concreteness a sudden transition, when the potential for
$\phi$ jumps from the negative values required for the contracting
phase to the positive ones required for the bounce. In the
transition, energy is conserved so that the raise in the potential
must be compensated by a negative contribution: $\dot\pi$ becomes
negative, which is indeed necessary for the bounce to happen. A
schematic representation of the potential is shown in Fig.
\ref{fig:potential}. $\Delta t$ represents the interval of time
during which the transition from the contracting to the bouncing
solution occurs. Notice that one can make $\Delta t$ much shorter
than $H^{-1}$, so that it can be treated as instantaneous for
the cosmological history, and at the same time much
longer than $M^{-1}$, so that the transition can be described
within the regime of the effective theory.

The sharp jump in the potential is rather unpleasant. However it is
important to stress that something so abrupt is unavoidable in a
bouncing cosmology where $H$ must evolve in a short time (of order
$H^{-1}$) from a large negative value before the bounce to a large positive value in the expanding phase.


Finally, as shown in Fig. \ref{fig:potential}, we imagine that at
the end of the bounce the energy associated with the ghost
condensate is converted to radiation, and the standard epoch of
cosmology begins.

\begin{figure}[th!!]
\psfrag{x}{$H<0$} \psfrag{y}{$H>0$} \psfrag{z}{$\dot{H}<0$}
\psfrag{t}{$\dot{H}>0$} \psfrag{u}{$H=0$} \psfrag{v}{$M_{\rm gc}^{-1}\ll
\Delta t\ll H^{-1}$} \psfrag{a}{$V(\phi)$} \psfrag{b}{$\phi$}
\psfrag{l}{$V(\phi)\sim -1/\phi^2$} \psfrag{m}{$V(\phi)\sim
\phi^2$}
\begin{center}
\includegraphics[scale=0.7]{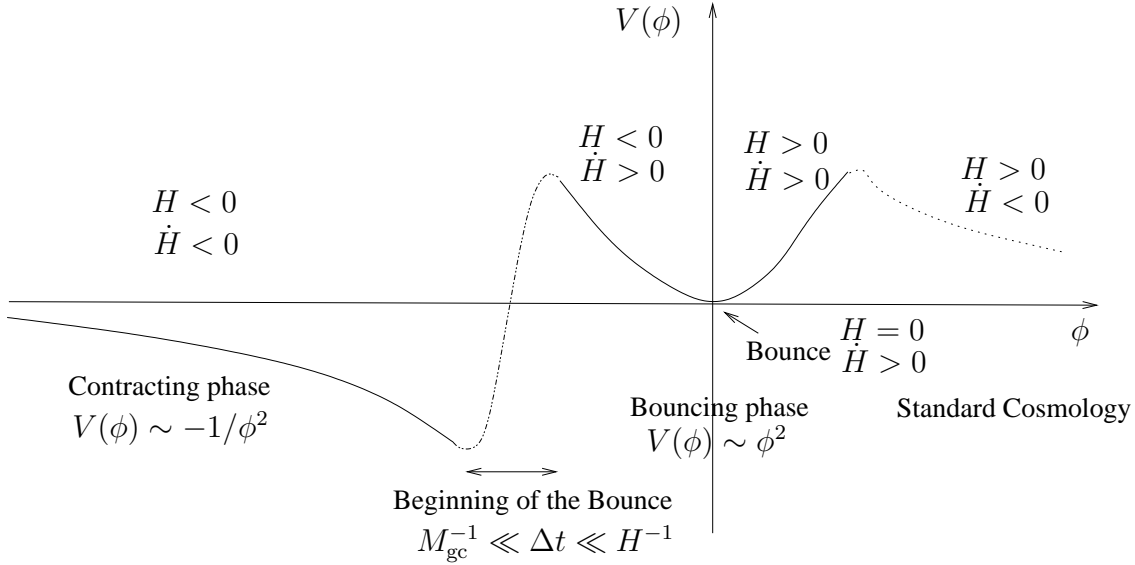}
\caption{\label{fig:potential} \small Schematic representation of
the potential of $\phi$ during the contracting and
the expanding phase.}
\end{center}
\end{figure}

We have not yet discussed the conversion of $\delta\psi$
perturbations to adiabatic ones. One can envisage many
possibilities. We assumed that when relevant modes freeze, the field
$\psi$ gives a sufficiently small contribution to the energy density
to neglect its mixing with gravity (see Appendix \ref{app: full
  action}). If this ceases to be true later on, a mixing between
$\delta\psi$ and $\zeta$ will occur.  Alternatively a mixing between $\phi$ and $\psi$ in the potential
can source the conversion.  In both cases an
approximately scale invariant spectrum for $\zeta$ is induced in a
way similar to what shown in Appendix \ref{app: full action} and this
will be preserved until the expanding phase: the general approach
advocated in \cite{Creminelli:2004jg} can in fact be applied and
explicitly checked in our smooth bouncing solution. Another
possibility is that the bouncing phase itself depends on the
isocurvature scalar $\psi$, a sort of ``variable bounce'' in analogy
with the variable decay scenario in inflation \cite{Dvali:2003em}. In this particular setting, one could imagine a
connection between the sudden transition in the potential and some
coupling between $\phi$ and $\psi$. A further possibility is that
$\delta\psi$ fluctuations are converted only in the expanding
phase, as in the curvaton \cite{Lyth:2001nq} and variable decay
models \cite{Dvali:2003em}.  Whatever is the mechanism, to find the final value of
$\zeta$ one has just
to solve the unperturbed history for different values of $\psi$ in the commonly used
``parallel Universes'' or $\delta N$ approach. This simplication
follows from the fact that $\delta\psi$ is relevant only when relevant
modes are out of the horizon.

\section{\label{sec:radiative}Naturalness of the exponential}
In inflation the origin of an approximately scale invariant spectrum of
perturbations can be traced back to the fact that the inflaton
potential is very flat, which implies that the Hubble
parameter is almost constant. In our scenario the origin of scale
invariance is completely different, as it comes from a precise
functional form of the potential $V(\psi)$, {\em i.e.}~a negative
exponential. It is worth emphasizing that while relevant
perturbations are produced, the potential of $\psi$ changes by many
orders of magnitude; an exponential is not a useful approximation to
simplify the algebra but a physical request: we need a negative exponential potential
over a large region.

One could object that this is not very satisfactory: in inflation we
just need a sufficiently shallow potential, while here we have to
require a precise functional form, which includes an infinite series of
non-renormalizable operators! \footnote{The situation resembles the
  one of inflationary models with a variation of the inflaton larger
  than the Planck scale. Without additional symmetries, we need a
  functional fine tuning to keep the potential sufficiently flat, {\em
  i.e.} we have to control an infinite series of non-renormalizable operators.}
We are obviously interested in the perturbative region of the
exponential, where the potential flattens out and the theory becomes
weakly coupled. The flatness of the potential in this region is
technically natural, in the sense that loop corrections will be
suppressed by derivatives of the potential, which are small. In other words
$\psi$ can be thought as an approximate Goldstone boson in this region.
However this does not help in selecting a specific function, among the
infinitely many which flatten out at infinity.

The point is that an exponential potential is {\em not} a generic infinite
series of non-renormalizable operators, but a particular one
which emerges in many examples. There are at least two possible origins of exponential
potentials. The first one is just a consequence of dimensional
reduction as it can be seen in this very simple example. If we take a
5-d theory compactified on $S^1$, the dimensional reduction of the
Einstein-Hilbert action gives at the level of zero modes
\be
S = \frac12 \int d^4x \sqrt{-g} \, M_5^3 \, T \, R \;,
\ee
where $g$ is the 4d metric and $R$ its Ricci scalar. The radion
field $T$ multiplies the 5d Planck scale $M_5$. If we do a conformal
rescaling to go to Einstein frame the radion gets a kinetic term of the
form
\be
\int d^4 x \sqrt{-g} \, M_4^2 \frac34 \, (\partial \log T)^2 \;,
\ee
where $M_4$ is the 4d Planck mass.
This implies that if the energy of the system depends polynomially on
the radius $T$, for example including a cosmological constant or just
at loop level because of the Casimir effect, the canonically
normalized radion will have an exponential potential, with a positive
or negative overall sign depending on the model. For instance a
negative cosmological constant induce a negative exponential
potential. The generalization of this trivial example leads to the
appearence of exponential potentials in supergravity compactifications
and therefore in string theory.

Another possible origin of exponential potentials in through
non-perturbative effects. In supersymmetric theories one can have
directions in field space which are flat in perturbation theory.
These can however be lifted by non-perturbative effects. For
instance if the flat direction enters in the gauge coupling,
instanton effects will give contribution going as powers of \be
e^{-\frac1{g^2}(\psi)} \;. \ee We thus see another possible origin
of an exponential potential.

We stress that, although an exponential shape of the potential is
rather ubiquitous in many scenarios, it is not clear whether the
condition $M \ll M_P$ and the request of a {\it negative} potential
can be fulfilled in string theory
\cite{Kallosh:2001ai,Khoury:2001iy} or in some other UV complete
theory.


\section{\label{sec:conclusions}Conclusions}
Inflation is a tremendously compelling scenario. However to make
progress it is crucial to have alternative models and to work out
their predictions, leaving the final word to experiments. It might be
dangerous to think that experiments are just measuring the parameters
of the only theory we have.

In this paper we described a bouncing cosmology, in which the bounce
is smooth and without pathologies. This is indeed possible in a theory with a
scalar with a modified kinetic term, like the
ghost condensate model with the addition of a potential.

What is new in this scenario with respect to previous bouncing
models, like the pre Big-Bang scenario and the ekpyrotic/cyclic one, is
that the physics of the bounce is explicit and under control. On the
contrary, in the other models one must {\em assume} that the bounce occurs and
its description must remain qualitative as it lies out of the regime of
validity of effective field theory and of the present understanding of
string theory. This is particularly important when one studies the evolution of
perturbations across the bounce, which is crucial to assess the
viability of a model.
To this smooth model of a bounce we add the mechanism to produce density
perturbations recently studied by Lehners, McFadden, Steinhardt and
Turok \cite{turok}
in the context of the ekpyrotic/cyclic model, that is an
isocurvature scalar moving along a negative exponential potential.

We therefore have an explicit and controllable model, in which predictions can be
derived: no observable gravitational waves, a high (but model
dependent) level of non-Gaussianity with a local shape and, most
importantly, a slightly blue spectral index. This model will be ruled
out if the present preliminary detection of a red tilt is confirmed by future data.

The model is clearly not very compelling as the Lagrangian of the
$\phi$ field seems really {\em ad hoc}.  This is the result of the fact
that it is difficult to construct a non-pathological system which induces a bounce. The stress
energy tensor must violate the null energy condition and this usually
leads to catasthrophic instabilities \cite{Creminelli:2006xe}. Maybe there are
much simpler systems leading to a bounce, either at the level of
effective field theory or in which quantum gravity is relevant. Notice however
that predictions depend only on the isocurvature scalar $\psi$ and
therefore are not sensitive to the explicit realization of the
contracting phase and of the bounce.

Alternatively, it might be that
the complicated Lagrangian for $\phi$ signals the fact that theories
which violate the null energy condition cannot be realized, in the
sense that our effective field theory cannot be UV completed. A deeper
understanding of the implications of the null energy condition is
clearly needed.

\section*{Acknowledgments}
It is a pleasure to thank Alberto Nicolis for collaboration at the
early stages of this work and Bobby Acharya, Nima Arkani-Hamed,
Sergei Dubovsky, Sameer Murthy, Andrea Romanino, Marco Serone,
Filippo Vernizzi, Giovanni Villadoro, Toby Wiseman and Matias
Zaldarriaga for useful discussions and comments.

\appendix

\section*{Appendix}
\section{Gravitational mixing between $\phi$ and $\psi$
\label{app: full action}}

In this appendix we provide the explicit Lagrangian for the system
of the Ghost Condensate $\phi$ plus the scalar field $\psi$ during
the contracting phase. With the resulting Lagrangian, we shall be
able to study the conditions under which we can neglect the
gravitational fluctuations in the $\delta\psi$ equation of motion,
justifying more regourosly what assumed in sec. \ref{sec:exp}.

In order to construct the action, we follow closely
\cite{Maldacena:2002vr}. We use ADM parametrization of the metric:
\begin{equation}
ds^2= -N^2 dt^2 + \hat g_{ij} (dx^i + N^i dt) (dx^j + N^j dt) \; ,
\end{equation}
where our background solution is of the form:
\begin{equation}
ds^2 = -dt^2 + a^2(t) d \vec{x}^2 \; ,
\end{equation}
where $a(t)\propto|t|^p$, and we consider general metric
fluctuations
\begin{equation}
N = 1 + \delta N \; , \quad N_{j} = \delta N_{j} \; , \quad
\hat{g}_{ij} = a^2(t) \delta_{ij} + \delta \hat{g}_{ij} \; .
\end{equation}
In this language, the Einstein-Hilbert action takes the form:
\begin{equation}\label{EH}
S_{\rm EH} = \frac 12 M_P^2 \int \! d^4 x \: \sqrt{-g} \, R =
\frac 12 M_P^2 \int \! d^3 x \, dt \: \sqrt{\hat g} \, \big[ N
R^{(3)} + \frac{1}{N} (E^{ij} E_{ij} - E^i{}_i {}^2) \big] \; ,
\end{equation}
where $R^{(3)}$ is the Ricci scalar of the
induced 3D metric. $E_{ij}$ is related to the extrinsic curvature
$K_{ij}$ of hypersurfaces of constant $t$, \be \label{extrinsic}
E_{ij} \equiv N K_{ij} = \frac 12 [{\partial_t {\hat g}}_{ij} -
\hat \nabla_i N_j - \hat \nabla_j N_i] \; , \ee where $\hat
\nabla$ is the covariant derivative associated to the induced 3D
metric $\hat g_{ij}$.

The ghost condensate $\phi$ and the scalar field $\psi$ has a
background solution of the form:
\begin{equation}
\phi(\vec{x},t)=\dot\phi_0\cdot
\left(t+\pi_0(t)+\delta\pi(\vec{x},t)\right) \; ,
\end{equation}
\begin{equation}
\psi(\vec{x},t)=\psi_0(t)+\delta\psi(\vec{x},t) \; .
\end{equation}
The action for the Ghost Condensate is given by\footnote{In
this Appendix we neglect effects due to higher derivative
terms that in the main text are proportional to $\bar M_{\rm gc}^2$.
Following \cite{Creminelli:2006xe}, one can
show that in order for a comoving mode $\zeta_k$ to be stable before
freezing ($\omega(k)\sim H$), $\bar M_{\rm gc}$ must be so small that
the $k^4$ term 
becomes irrelevant well before this time. Furthermore, as discussed in
the main text, the Jeans instability that $\bar
M_{\rm gc}$ would induce is subdominant with respect to the
$k^2$ term proportional to $\dot H$. Here we are interested
in studying the gravitational mixing between $\phi$ and $\psi$,
and this occurs only around or after the freezing time, so that we
can safely neglect the higher derivative terms in our discussion. }:
\begin{equation} \label{actionGC}
S_{\rm GC} = \int \! d^4 x \: \sqrt{-g} \, P(X,\phi)=\int \! d^3 x
\, dt \: \sqrt{\hat g} N \, P(X,\phi) \; ,
\end{equation}
where $X=- g^{\mu\nu}\partial_\mu\phi\partial_\nu\phi$. The action for the field
$\psi$ is given by:
\begin{equation} \label{actionpsi}
S_{\psi}
=\frac{1}{2}\int \! d^3 x \, dt \: \sqrt{\hat g} \; \big[N^{-1}
\left(\dot\psi-N^i\partial_i\psi\right)^2-N
\hat{g}^{ij}\partial_i\psi\partial_j\psi-2N V(\psi)\big]\; ,
\end{equation}
where $V(\psi)=V_0 e^{\psi/M}$. We see that $\psi$ and $\phi$ mix
only through gravity. As we said in the main part of the paper, at
a certain point during the contracting phase, or after the bounce,
there must be some direct interaction between the $\phi$ and $\psi$ to
allow for the conversion of the isocurvature fluctuations into
adiabatic one. This however must occur at least after all the
relevant cosmological modes have exit the horizon. This is a much
later time with respect to the one we are considering here. For this
reason, here we can neglect all direct interactions between $\psi$
and $\phi$.

The ADM formalism is designed so that one can think of
$\hat{g}_{ij},\, \phi$ and $\psi$ as dynamical variables, and $N$
and $N^i$ as Lagrange multipliers. We will choose a gauge for
$g_{ij}, \phi$ and $\psi$ that will fix time and spatial
reparametrizations. We find convenient to define the gauge:
\begin{equation}
\delta\pi=0, \quad \hat{g}_{ij}=a^2(t)(1+2\zeta)\delta_{ij} \ .
\end{equation}
This gauge fixes completely time and space diffeomorphisms at non
zero momentum. It represent a gauge where the ghost condensate is
uniform, and taken as the time variable. The dynamical
degrees of freedom are $\zeta$ and $\delta\psi$. In the limit
where the gravitational mixing of $\psi$ with $\phi$ is
negligible, we can think of $\zeta$ as the scalar degree of
freedom associated with the ghost condensate. In this gauge, the
action for $\phi$ takes the form:
\begin{eqnarray}\label{GCtadpole}
S_{\rm GC}=\int \! d^3 x \, dt \: \sqrt{\hat g} \;
\Big[\frac{\left(-2\,M^2 + M^2_P\;p\right)}{t^2}
\frac{1}{N}+\frac{\left( 1 - 3\,p \right) \,\left( -2\,M^2 +
M^2_P\,p \right) }{t^2}
N+\frac{1}{2}\frac{\tilde{M_{\rm gc}}^2}{t^2}(\delta N)^2\Big]
\end{eqnarray}
The tadpole terms in (\ref{GCtadpole}) are chosen in order to
ensure that the background solution has exactly the form
$a(t)\propto|t|^p$, and
$\dot\psi_0(t)=-\frac{2M}{t},V(\psi_0(t))=-\frac{2M^2(1-3p)}{t^2}$
as the one we used in the main part of the paper. This means that
the form of the function $P(X,\phi)$ used here and the
corresponding $\phi_0(t)$ solution will be slightly different (by
order $p$ corrections) from the one used in the main part. Thought
the physical implications do not change, considering an exact
solution simplifies relevantly the study of the action of the
system.

To find the action for $\zeta$ and $\delta\psi$, we solve for $N$
and $N^i$ through their equations of motion and plug the result
back in the action. This procedure is allowed because $N$ and
$N^i$ are Lagrange multipliers. The equations of motion for $N$
and $N^i$ read at first order:
\begin{eqnarray}
&&\partial_i\Big[ \frac{2\, M}{t}\delta\psi + 2\,\MP^2 H(t)\,
\delta N - 2\,\MP^2\dot\zeta  \Big]=0 \, , \\ \nonumber &&
\frac{2\,M^2_Pp\,( 1 - 3\,p ) +\tilde{M_{\rm gc}}^2}{t^2} \, \delta N +
\frac{2 M( 1 - 3\,p )}{t^2}\delta\psi +
         \frac{2 M}{t}\dot{\delta\psi} -
           \frac{ 2\,M^2_P\,p\,}{t} \chi  - \frac{ 6\,M^2_P\,p\,}{t}\dot\zeta   -
      2\, M^2_P\frac{\partial^2}{a^2}\zeta=0\; ,
\end{eqnarray}
where for convenience we have defined:
\begin{equation}
N^i=\partial_i\psi\, , \quad \partial^2\psi=\chi \, .
\end{equation}
We can solve these equations to first order to obtain:
\begin{eqnarray}
&&N_1=- \frac{M}{\MP^2\,p}\delta\psi +
  \frac{t}{p}\dot\zeta \, ,\\ \nonumber
&&\chi=-\frac{ M\,\tilde M_{\rm gc}^2}{2 \MP^4\,p^2\,t} \delta\psi+
  \frac{M\,}{\MP^2\,p}\dot{\delta\psi} + \left(\frac{1}{p} +
  \frac{\tilde M_{\rm gc}^2}{2\,\MP^2\,p^2}\right)\dot\zeta -
  \frac{t}{p\,a^2}(\partial_i^2\zeta)\, .
\end{eqnarray}
In order to find the quadratic action for $\zeta$ and
$\delta\psi$, we do not need the second order solutions for $N$
and $N^i$. The reason for this is that the second order term in
$N$ will multiply the constraint $\frac{\partial {\cal
L}}{\partial  N}$ evaluated at zeroth order, which vanishes since
the zeroth order solution obeys the equations of motion. A similar
argument holds for $N^i$.

Substituting in the action $S=S_{{\rm EH}}+S_{{\rm
GC}}+S_{\delta\psi}$, after performing some integration by parts,
we find an action of the form:
\begin{equation}
S=\int \! d^3 x \, dt \:a^3 \big[{\cal L}_{\zeta}+{\cal
L}_{\psi}+{\cal L}_{{\rm Mixing}}\big] \; ,
\end{equation}
where:
\begin{eqnarray}
&&{\cal L}_{\zeta}=\frac{(2 M^2_Pp
+\tilde{M_{\rm gc}}^2)}{2p^2}\dot\zeta^2-\frac{\MP^2}{
p}\left(\frac{\partial_i}{a}\zeta\right)^2 \; , \\ \nonumber &&
{\cal
L}_{\delta\psi}=\frac{1}{2}\dot{\delta\psi}^2-\frac{1}{2}\left(\frac{\partial_i}{a}\delta\psi\right)^2+
\left(1-3p-\frac{6 M^2}{\MP^2}+\frac{M^2\tilde{M_{\rm gc}}^2}{2\MP^4
p^2}\right)\frac{1}{t^2}\delta\psi^2 \; , \\ \nonumber &&{\cal
L}_{{\rm Mixing}}=-\frac{M \tilde{M_{\rm gc}}^2}{\MP^2 p^2 t}\dot\zeta
\delta\psi+2\frac{M}{p}\dot\zeta
\dot{\delta\psi}+2\frac{\MP}{p^2}\left(\frac{\partial_i^2}{a^2}\zeta\right)\delta\psi
\; .
\end{eqnarray}
We see that, neglecting the mixing, the speed of sound of $\zeta$
is given by:
\begin{equation}
c_s^2=\frac{2 p \MP^2}{(2\MP^2 p+\tilde M_{\rm gc}^2)}\simeq \frac{2 p
\MP^2}{\tilde M_{\rm gc}^2}\ll 1 \; ,
\end{equation}
where in the second and third passage we have used the fact that
for the validity of the ghost condensate effective theory we need
to have:
\begin{equation}
\frac{ p \MP^2}{\tilde M_{\rm gc}^2}\ll 1 \; .
\end{equation}
The speed of sound of $\zeta$ must be very small. It is useful to
write the former Lagrangians in terms of $c_{s}$. Keeping only
leading order terms in $c_s$, we obtain:
\begin{eqnarray}
&&{\cal L}_{\zeta}=\frac{\MP^2}{2 p\; c_{s}^2
}(\dot\zeta^2-c_s^2\left(\frac{\partial_i}{a}\zeta\right)^2) \; ,
\\ \nonumber &&{\cal
L}_{\delta\psi}=\frac{1}{2}\dot{\delta\psi}^2-\frac{1}{2}\left(\frac{\partial_i}{a}\delta\psi\right)^2+
\left(1-3p-\frac{6 M^2}{\MP^2}+\frac{M^2}{2\MP^2p\;
c_s^2}\right)\frac{1}{t^2}\delta\psi^2 \; , \\ \nonumber &&{\cal
L}_{{\rm Mixing}}=-2\frac{M}{p\; c_s^2 t}\dot\zeta
\delta\psi+2\frac{M}{p}\dot\zeta
\dot{\delta\psi}+2\frac{\MP}{p^2}\left(\frac{\partial_i^2}{a^2}\zeta\right)\delta\psi
\; .
\end{eqnarray}
If we look at the Lagrangian for $\delta\psi$, we see that the
inclusion of gravitational perturbations has produced two effects:
the appearance of a direct coupling between $\zeta$ and $\psi$,
and a correction to the mass term which is generated when we plug
back into ${\cal L}_{\delta\psi}$ the solutions for $N$ and $N^i$.
Obviously, the form of the mixing terms does depend on the nature
of the field which leads the contraction. However, there is
a regime where $\psi$ is so subdominant that these mixing effects
are negligible. In this case, the nature of the field which drives
the contraction is irrelevant for most of the results. This is the
case in which we concentrated in the paper. Now we are able to
explicitly check what is the condition under which this is
verified. The condition will turn out to be not exactly
$\rho_\psi/\rho_{\rm tot}\ll 1$, because we shall have to impose
the mixing to be so small not to alter the calculation we did for
the tilt, and also because the unmixed $\zeta$ has a small speed
of sound. It is rather straightforward, comparing for example the
terms of mixing with the diagonal terms for $\delta\psi$, that the
effects of gravitational mixing are negligible for the
$\delta\psi$ equation of motion if
\begin{equation}\label{unmixingcondition}
\frac{\rho_\psi}{\rho_{\rm tot}}=\frac{M^2}{p \MP^2}\ll {\rm
Min}\{p^2 c_s^{-1},p\; c_s^2\}
\end{equation}
It is interesting to note that even though this condition is
satisfied, which means that $\zeta$ has no effect on $\delta\psi$,
still at late times, $t\gg 1/k$, the effect of $\delta\psi$ on
$\zeta$ is not negligible. This is just the effect that on large scales,  after
the particular mode has freezed, different values of $\delta\psi$ induce small differences
in the expansion of the separate regions. The induced $\zeta$ is
scale invariant and of size:
\begin{equation}
\zeta_{\rm induced}\sim \frac{M}{\MP^2 t}\lesssim
\frac{1}{100}\frac{M^2}{\MP^2}\ll \frac{p^2 c_s^2}{100}\ll 10^{-5}
\, ,
\end{equation}
where the first inequality comes from the limit on
non-gaussinities, the second from (\ref{unmixingcondition}), and
the third from the fact that, because of the constraint on the
tilt of the two point function, $p\lesssim {\rm few}\times
10^{-2}$, and $c_s\ll 1$ for the validity of the effective field
theory. The induced $\zeta$ is therefore too small, justifying our
request for a separate mechanism for conversion of isocurvature
perturbations into adiabatic ones.

An alternative interesting regime, different from the one on
which we concentrated in this paper, is the one where the inequality
(\ref{unmixingcondition}) begins to be violated. In this case, two
important things might occur. On the one hand, it is clear from
the mass term in ${\cal L}_{\delta\psi}$ that the factors of $p$ that determine
the tilt might be overcome, inducing possibly a red tilt. On the other hand, the
induced $\zeta$ might become large enough, without loosing its scale invariance,
so that there might be no need for an explicit conversion mechanism
between isocurvature and adiabatic perturbations. Though these
results would be appealing, the predictions would depend also on
the details of the field that drives the contracting phase, and
therefore they would be much more model dependent. For this
reason, in the main part of the paper we concentrated in the region where the
condition (\ref{unmixingcondition}) is satisfied and all the effects coming from
mixing with gravity can be safely neglected.

\end{document}